\documentclass[journal]{IEEEtran}
\usepackage{amsmath} \allowdisplaybreaks
 \usepackage{changebar}
\usepackage{amssymb}

\usepackage[dvips]{graphicx}

\ifCLASSOPTIONcompsoc
\usepackage[caption=false,font=normalsize,labelfont=sf,textfont=sf]{subfig}
\else
\usepackage[caption=false,font=footnotesize]{subfig}
\fi

\usepackage{array}
\usepackage{multirow}  
\usepackage{array} 
\newcommand{\PreserveBackslash}[1]{\let\temp=\\#1\let\\=\temp}
\newcolumntype{C}[1]{>{\PreserveBackslash\centering}p{#1}}
\newcolumntype{R}[1]{>{\PreserveBackslash\raggedleft}p{#1}}
\newcolumntype{L}[1]{>{\PreserveBackslash\raggedright}p{#1}}
\usepackage{chemarrow}
\usepackage{mdwmath}
\usepackage{stfloats}

\usepackage[lined,boxed,commentsnumbered, ruled]{algorithm2e}

\begin{document}
%
\title{A Class of Low-Interference \emph{N}-Continuous OFDM Schemes}
%
%
%

\author{Peng~Wei, Lilin~Dan, Yue~Xiao, Wei~Xiang, and Shaoqian~Li
\thanks{The authors are with the school of National Key Laboratory of Science and Technology on Communications, University of Electronic Science and Technology of China, Chengdu, China (e-mail: wpwwwhttp@163.com; \{lilindan, xiaoyue\}@uestc.edu.cn). Wei Xiang is with the school of Mechanical and Electrical Engineering Faculty of Health, Engineering and Sciences, University of Southern Queensland, Austrialia (e-mail: wei.xiang@usq.edu.au).}}%
\maketitle

\begin{abstract}
\emph{N}-continuous orthogonal frequency division multiplexing (NC-OFDM) was demonstrated to provide significant sidelobe suppression for baseband OFDM signals. However, it will introduce severe interference to the transmit signals. Hence in this letter, we specifically design a class of low-interference NC-OFDM schemes for alleviating the introduced interference. Meanwhile, we also obtain an asymptotic spectrum analysis by a closed-form expression. It is shown that the proposed scheme is capable of reducing the interference to a negligible level, and hence to save the high complexity of signal recovery at the receiver, while maintaining similar sidelobe suppression performance compared to traditional NC-OFDM.
\end{abstract}


\IEEEpeerreviewmaketitle

\section{Introduction}

\IEEEPARstart{O}{rthogonal} frequency division multiplexing (OFDM) has been widely adopted in wireless communications 
due to its high-speed data transmission and inherent robustness against the inter-symbol interference (ISI). However, traditional rectangularly pulsed OFDM exhibits large spectral sidelobes, resulting in severe out-of-band power leakage. In this case, traditional OFDM will introduce high interference to the adjacent channels.

For alleviating this problem, various methods have been proposed for sidelobe suppression in OFDM \cite{Ref3}-\cite{Ref14}. More specifically, the windowing technique \cite{Ref3} extends the guard interval at the cost of spectral efficiency reduction. Cancellation carriers \cite{Ref4} 
will consume extra power and incur a signal-to-noise ratio (SNR) loss. Precoding methods \cite{Ref6}-\cite{Ref8} require complicate decoding processing to eliminate the interference. 

Against this background, NC-OFDM \cite{Ref9} is a class of efficient sidelobe suppression techniques by making the OFDM signal and its first \emph{N} derivatives continuous. However, a disadvantage of NC-OFDM is the high implementation complexity in the  transmitter. For reducing the complexity, some simplified schemes have been proposed in \cite{Ref10} and \cite{Ref11}. In \cite{Ref15}, time-domain NC-OFDM (TD-NC-OFDM) was proposed for reducing the complexity of NC-OFDM, by transforming traditional frequency-domain processing into the time domain. On the other hand, traditional NC-OFDM will introduce severe interference so as to degrade the bit error rate (BER) performance. For alleviating this problem, \emph{N}-continuous symbol padding OFDM (NCSP-OFDM) was recently developed in \cite{Ref14} at the cost of increased complexity in the transmitter. Meanwhile, to enable low-complexity signal recovery in NC-OFDM, several techniques \cite{Ref12}-\cite{Ref13} have been proposed. However, those techniques will also result in complexity load or efficiency reduction.  

For reducing the interference of NC-OFDM and avoiding complex signal recovery at the receiver, this letter proposes a low-interference NC-OFDM scheme by adding an improved smooth signal in the time domain. Firstly, in the proposed scheme, the smooth signal is generated in a novel way as the linear combination of designed basis signals related to rectangular pulse. Secondly, the basis signals are smoothly truncated by a preset window function to obtain the short-duration smooth signal with shorter length than a cyclic-prefixed OFDM symbol and low interference. Thirdly, the short-duration smooth signal is overlapped onto only part of the OFDM symbol to reduce the interference. Furthermore, we give an asymptotic expression of the spectrum feature of the low-interference NC-OFDM signal, to show that the proposed scheme can maintain the similar sidelobe suppression to traditional NC-OFDM. Lastly, the complexity analysis denotes that the proposed scheme can significantly reduce the complexity of NC-OFDM.

\section{\emph{N}-continuous OFDM}

In conventional NC-OFDM \cite{Ref9}, the \emph{i}th transmit symbol $\bar{y}_i(t)$ follows 
\begin{equation}
  \bar{y}_i(t)=\sum\limits^{K-1}_{r=0}{\bar{x}_{i,k_r}e^{j2\pi k_r\Delta ft}}, -T_{\rm cp}\leq t <T_{\rm s}
  \label{Eqn1}
\end{equation}
and
\begin{equation}
  \left. \bar{y}^{(n)}_i(t)\right|_{t=-T_{\rm cp}}=\left. \bar{y}^{(n)}_{i-1}(t)\right|_{t=T_{\rm s}},
  \label{Eqn2}
\end{equation}
where $\bar{x}_{i,k_r}$ is the precoded symbol on the \emph{r}th subcarrier with the subcarrier index set $\mathcal{K}=\{k_0, k_1, \ldots, k_{K-1}\}$, $y^{(n)}_{i}(t)$ is the \emph{n}th-order derivative of $y_{i}(t)$ with $n\in\mathcal{U}_N \triangleq \{0,1,\ldots,N\}$, the frequency spacing is $\Delta f=1/T_{\rm s}$, $T_{\rm s}$ is the symbol duration, and $T_{\rm cp}$ is the cyclic prefix (CP) duration. 

For satisfying Eq. \eqref{Eqn2}, the \emph{N}-continuous processing can be summarized as \cite{Ref9}
\begin{equation}
\begin{cases}
 \bar{\mathbf{x}}_i=\mathbf{x}_0, & i=0  \\
 \bar{\mathbf{x}}_i=(\mathbf{I}_K-\mathbf{P})\mathbf{x}_i+\mathbf{P}\mathbf{\Phi}^H\bar{\mathbf{x}}_{i-1}, & i>0,
\end{cases}
  \label{Eqn3}
\end{equation}
where ${\mathbf{x}}_i\!\!=\!\![{x}_{i,k_0},\ldots,{x}_{i,k_{K-1}}]^T$, $\bar{\mathbf{x}}_i\!\!=\!\![\bar{x}_{i,k_0},\ldots,\bar{x}_{i,k_{K-1}}]^T$, $\mathbf{I}_K$ is the identity matrix, $\mathbf{P}=\mathbf{\Phi}^H\mathbf{A}^H(\mathbf{A}\mathbf{A}^H)^{-1}\mathbf{A}\mathbf{\Phi}$, $\mathbf{\Phi}\triangleq {\rm diag}(e^{j\varphi k_0},e^{j\varphi k_1},\ldots,e^{j\varphi k_{K-1}})$, $\varphi=-2\pi\beta$ with $\beta=T_{\rm cp}/T_{\rm s}$, and 
$$\mathbf{A}=\begin{bmatrix} 1 & 1 & \cdots & 1 \\ k_0 & k_1 & \cdots & k_{K-1} \\ \vdots & \vdots & {} & \vdots \\ k^{N}_{0} & k^{N}_{1} & \cdots & k^{N}_{K-1} \end{bmatrix}.$$

\section{Low-Interference NC-OFDM Scheme}

TD-NC-OFDM is recently proposed in \cite{Ref15} for transforming the processing of NC-OFDM into the time domain. In TD-NC-OFDM, the smooth signal is added with the time sampling interval $T_{\rm samp}=T_{\rm s}/M$ where \emph{M} is the length of the symbol. Following the basic idea of TD-NC-OFDM, the proposed low-interference scheme also adds a smooth signal $w_i(m)$ onto the OFDM signal $y_i(m)$ in the time domain, given as
\begin{equation}
  \bar{y}_i(m)=y_i(m)+w_i(m),
  \label{Eqn4}
\end{equation}
where $m\in\mathcal{M}=\{-M_{\rm cp}, -M_{\rm cp}+1, \ldots, M-1\}$ and $M_{\rm cp}$ is the length of a CP.
Thus, to make the OFDM signal \emph{N}-continuous, $w_i(m)$ should satisfy 
\begin{equation}
  \left.{w}^{(n)}_i\!(m)\right|_{m=-M_{\rm cp}}\!\!=\left.y^{(n)}_{i-1}\!(m)\right|_{m=M}
  \!-\left.y^{(n)}_{i}\!(m)\right|_{m=-M_{\rm cp}}\!.
  \label{Eqn5}
\end{equation}


To construct $w_i(m)$ with low interference, $w_i(m)$ is just added in the front part of each CP-prefixed OFDM symbol. Thus, the proposed linear-combination design of $w_i(m)$ is described as
\begin{equation}
  {w}_i(m)=\left\{\begin{matrix}
          \sum\limits^{N}_{n=0}{{b}_{i,n}\tilde{f}_n(m)}, & m\in \mathcal{L} \\
          0, & m\in \mathcal{M}\backslash \mathcal{L},
\end{matrix}\right.
  \label{Eqn6}
\end{equation}
where $\mathcal{L}\triangleq\left\{-M_{\rm cp},-M_{\rm cp}+1,\ldots,-M_{\rm cp}+L-1\right\}$ indicates the location of $w_i(m)$ with length \emph{L}, and the basis signals $\tilde{f}_n(m)$ belong to the basis set $\mathcal{Q}$, defined as
\begin{align}
  {\mathcal{Q}}&\triangleq\bigg\{{\mathbf{q}}_{\tilde{n}}\left|{\mathbf{q}}_{\tilde{n}}
  =\Big[\tilde{f}_{\tilde{n}}(-M_{\rm cp}),\tilde{f}_{\tilde{n}}(-M_{\rm cp}+1),\ldots, \right.   \nonumber \\
 & \qquad\qquad\qquad \tilde{f}_{\tilde{n}}(-M_{\rm cp}+L-1)\Big]^T, \tilde{n}\in \mathcal{U}_{2N}\bigg\},
  \label{Eqn7}
  \end{align}
where $\mathcal{U}_{2N}=\{0,1,\ldots,2N\}$. For achieving the low-interference smooth signal $w_i(m)$ in Eq. \eqref{Eqn6}, the design of the basis signals $\tilde{f}_{\tilde{n}}(m)$ and the linear combination coefficients $b_{i,n}\in\mathbf{b}_i=[b_{i,0},b_{i,1},\ldots,b_{i,N}]^T$ will be specified as follows.

On the one hand, the time duration of $\tilde{f}_{\tilde{n}}(m)$ is truncated by a preset window function $g(m)$, which is considered as a smooth and zero-edged window function, such as triangular, Hanning, or Blackman window function. Then, the truncated basis signals can be given by
\begin{equation}
\tilde{f}_{\tilde{n}}(m)=\left\{\begin{matrix}
          f^{(\tilde{n})}(m)g(m)u(m), & m\in\mathcal{L}  \\
          0, & m\in\mathcal{M}\backslash \mathcal{L},
\end{matrix}\right.
  \label{Eqn8}
\end{equation}
where $u(m)$ denotes the unit-step function, and $f^{(\tilde{n})}(m)$ is calculated by the rectangularly OFDM pulse \cite{Ref15}, given as
\begin{equation}
f^{(\tilde{n})}(m)= 1/M\left(j2\pi/M\right)^{\tilde{n}}\sum\limits_{k_r\in\mathcal{K}}{k^{\tilde{n}}_re^{j\varphi k_r}e^{j2\pi\frac{k_r}{M}m}}.
  \label{Eqn9}
\end{equation}

On the other hand, by substituting Eqs. \eqref{Eqn6}-\eqref{Eqn9} into Eq. \eqref{Eqn5}, the coefficients $b_{i,n}$ can be calculated as
\begin{equation}
  {\mathbf{b}}_i=\mathbf{P}^{-1}_{\tilde{f}}( {\mathbf{P}}_1\mathbf{x}_{i-1}-\mathbf{P}_2\mathbf{x}_i),
  \label{Eqn10}
\end{equation}
where $\mathbf{P}_{\tilde{f}}$  is a $(N+1)\times(N+1)$ symmetric matrix, given as 
$$\mathbf{P}_{\tilde{f}}=\begin{bmatrix}
\tilde{f}^{(0)}(-M_{\rm cp}) &  \cdots &  \tilde{f}^{(N)}(-M_{\rm cp}) \\ 
\tilde{f}^{(1)}(-M_{\rm cp}) & \cdots &  \tilde{f}^{(N+1)}(-M_{\rm cp}) \\ 
\vdots  & {} &\vdots\\
\tilde{f}^{(N)}(-M_{\rm cp})  & \cdots &  \tilde{f}^{(2N)}(-M_{\rm cp}) \\ 
\end{bmatrix},$$
$${\mathbf{P}}_1=\frac{1}{M}\begin{bmatrix}
1  & \cdots &  1 \\ 
{j2\pi k_0}/{M}  & \cdots &  {j2\pi k_{K-1}}/{M} \\ 
\vdots  & {} &\vdots\\
\left({j2\pi k_0}/{M}\right)^{N}  & \cdots &  \left({j2\pi k_{K-1}}/{M}\right)^{N} \\ 
\end{bmatrix},$$
and $\mathbf{P}_{2}={\mathbf{P}}_1\mathbf{\Phi}$.

Finally, $w_i(m)$ is added onto only part of the CP-inserted OFDM symbol to achieve the \emph{N}-continuous symbol $\bar{\mathbf{y}}_i$, as
\begin{equation}
  \bar{\mathbf{y}}_i=\left\{\begin{matrix}
      \mathbf{y}_i+\left[\begin{array}{c}\mathbf{Q}_{\tilde{f}}{\mathbf{b}}_i \\ \mathbf{0}_{(M-L)\times 1}\end{array}\right] & 0\leq i\leq M_{\rm s}-1, \\
      \mathbf{Q}_{\tilde{f}}{\mathbf{b}}_i &  i=M_{\rm s},
\end{matrix}\right.
  \label{Eqn11}
\end{equation}
where $\mathbf{Q}_{\tilde{f}}=\{\mathbf{q}_0,\mathbf{q}_1,\ldots,\mathbf{q}_N\}$, $M_{\rm s}$ is the number of the OFDM symbols, and $\mathbf{x}_{-1}$ is initialized as $\mathbf{x}_{-1}=\mathbf{0}_{K\times 1}$ since the back edge of $\mathbf{y}_{-1}$ equals to zero.

In general, the proposed low-interference NC-OFDM scheme is illustrated as follows.
\begin{algorithm}
\NoCaptionOfAlgo
\caption{ {\bfseries Low-Interference NC-OFDM Algorithm}}
\BlankLine
\SetAlgoNoLine
\LinesNumbered
\KwIn{$\mathbf{x}_{i}$, $\mathbf{y}_{i}$ with CP}
\KwOut{$\bar{\mathbf{y}}_i$}
\BlankLine
Initialization : $\mathbf{x}_{-1}=\mathbf{0}_{K\times 1}$, $i=0$, $\mathbf{Q}_{\tilde{f}}$, $\mathbf{P}^{-1}_{\tilde{f}}$,  $\mathbf{P}_{1}$, $\mathbf{P}_{2}$;\\
\While{$i\le M_{\rm s}-1$}
{
Calculate the coefficients $b_{i,n}$ by Eq. \eqref{Eqn10};\\
Generate the smooth signal $\mathbf{w}_i=\mathbf{Q}_{\tilde{f}}{\mathbf{b}}_i$;\\
Add $\mathbf{w}_i$ onto the OFDM signal by Eq. \eqref{Eqn11}; \\
{$i=i+1$;}
}
\eIf{$i=M_{\rm s}$}
{$\mathbf{w}_i$ is appended to the transmit signal as Eq. \eqref{Eqn11};}
{Terminate.}
\end{algorithm}

According to the above processing, the main advantage of the improved smooth signal $w_i(m)$ is that its length has been effectively truncated to \emph{L}, by a careful design in Eq. \eqref{Eqn8}. Furthermore, for $L<M+M_{\rm cp}$, only parts of the OFDM signal are overlapped with the interference $w_i(m)$. More especially, since the front part of the OFDM symbol is CP, the interference to the real data is further mitigated. In Section V, we will show that the influence of $w_i(m)$ has been effectively reduced. 


%
%

\section{Theoretical Analysis of Spectrum and Complexity}

\subsection{Spectral Analysis}
 
Assume that the first \emph{N}-1 derivatives of the smoothed OFDM signal are continuous, and the \emph{N}th-order derivative has finite amplitude discontinuity \cite{Ref18}. Thus, according to the definition of the power spectrum density (PSD) \cite{Ref8} and the relationship between spectral roll-off and continuity in \cite{Ref16}, the PSD of the low-interference NC-OFDM signal is expressed by 
 \begin{align}
 {\Psi}(f) 
    \!=\! \lim\limits_{U\rightarrow \infty}\!{\frac{1}{2UT}E\!\!\left\{\!\left|\sum\limits^{U-1}_{i=-U}{\!\!\frac{\mathcal{F}
   \left\{\bar{y}^{(N)}_{i}(t)\right\}}{\left(j2\pi f\right)^N}e^{-j2\pi fiT}}\right|^2\!\right\}},
   \label{Eqn12}
 \end{align}
where $T=T_{\rm s}+T_{\rm cp}$.

Eq. \eqref{Eqn12} indicates that the spectrum of the low-interference NC-OFDM signal is related to the expectation of $\bar{y}^{(N)}_i(t)$ multiplied by $f^{-N}$. In this letter, the conventional Blackman window function is used as an example, given as $g(t)=0.42-0.5\cos{(2\pi \rho t)}+0.08\cos{(4\pi \rho t)}$ where $\rho=1/\left((2L-2)T_{\rm samp}\right)$. By substituting Eqs. \eqref{Eqn5} \eqref{Eqn6} \eqref{Eqn8} and \eqref{Eqn9} into Eq. \eqref{Eqn12}, the PSD of the smoothed OFDM signal is expressed by
  \begin{align}
  {\Psi}(f)
        &=\lim\limits_{U \rightarrow \infty}\frac{1}{2UT}E\Bigg\{\bigg|\sum\limits^{U-1}_{i=-U}e^{-j2\pi fiT}\left(fT_{\rm s}\right)^{-N}   \nonumber\\
        & \cdot \sum\limits_{k_r\in \mathcal{K}}\!\!{k^N_rx_{i,k_r}\mathrm{sinc}\left(f_r(1+\beta)\right)e^{j\pi f_r(1-\beta)}} \!\!+\!\frac{1}{T}\!\sum\limits^{N}_{n=0}{b}_{i,n}   \nonumber\\
        & \cdot \sum\limits^{N}_{\bar{n}=0}\!\!\left(\begin{matrix}
            N \\ \bar{n} \end{matrix}\right)\!\!\left(\!{j2\pi}/{T_{\rm s}}\!\right)^{n-\bar{n}} 
         \!\!\! \sum\limits_{k_r\in \mathcal{K}}\!{k^{N-\bar{n}+n}_rG_{\bar{n}}(f)}\bigg|^2\!\Bigg\},
   \label{Eqn13}
 \end{align}
where $\text{sinc}(x)\triangleq \sin(\pi x)/(\pi x)$, $f_r=k_r-T_{\rm s}f$, $\left(\begin{matrix} N \\ \bar{n} \end{matrix}\right)$ is the binomial coefficient,
 \begin{eqnarray*}
  \!\!\!&\!\!\! G_{\bar{n}}(f)=e^{j\tilde{f}_r}\Bigg(0.42^{(\bar{n})}\dfrac{\sin(\pi \mu f_r)}{\pi f_r/T_{\rm s}}-\dfrac{0.5(2\pi\rho)^{\bar{n}}\cos{\left(\pi\mu f_r\right)}}{1-\left(\rho T_{\rm s}/f_r\right)^2} \\
  \!\!\! &\!\!\! \cdot \left(\dfrac{\cos{\left(\pi\bar{n}/2\right)}}{j\pi f_r/T_{\rm s}}-\pi\rho\dfrac{\sin{\left(\pi\bar{n}/2\right)}}{\left(\pi f_r/T_{\rm s}\right)^2}\right) \\
   \!\!\!&\!\!\! + \dfrac{0.08(4\pi\rho)^{\bar{n}}\sin{\left(\pi\mu f_r\right)}}{1-\left(2\rho T_{\rm s}/f_r\right)^2}
    \!\left(\!\!\dfrac{\cos{\left(\pi\bar{n}/2\right)}}{\pi f_r/T_{\rm s}}\!-j2\pi\rho\dfrac{\sin{\left(\pi\bar{n}/2\right)}}{\left(\pi f_r/T_{\rm s}\right)^2}\!\right)\!\!\!\Bigg),
\end{eqnarray*}
 $\tilde{f}_r\!=\!\pi f_r(T_{\rm p}-2T_{\rm cp})/T_{\rm s}$ with $T_{\rm p}\!=\!1/(2\rho)$, and $\mu\!=\!T_{\rm p}/T_{\rm s}$. 

\begin{figure}[h]
\centering
\includegraphics[width=3.5in]{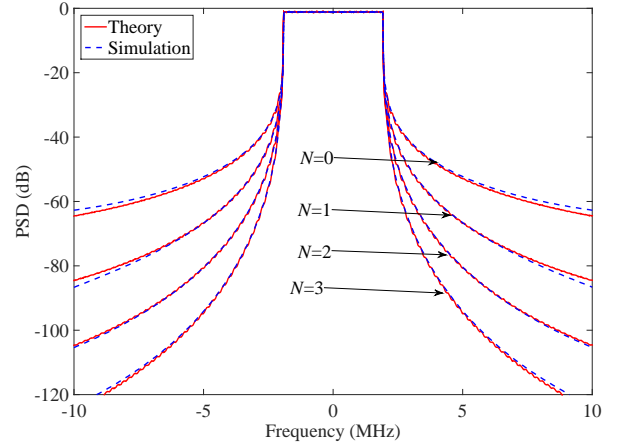}
\DeclareGraphicsExtensions.
\caption{PSD comparison between the analytical and simulation results in low-interference NC-OFDM.}
  \label{Fig2}
\end{figure}

Eq. \eqref{Eqn13} shows that the power spectral roll-off of the smoothed signal, whose first \emph{N}-1 derivatives are continuous, decays with $f^{-2N-2}$. Moreover, the expression of $G_{\bar{n}}(f)$ reveals that the sidelobe is affected by the length of $w_i(t)$, so that the selection of \emph{L} is important to balance BER and sidelobe suppression performance. Fig. \ref{Fig2} compares the theoretical and simulation results of the low-interference scheme with \emph{L}=$M_{\rm cp}$=144, \emph{M}=2048, where the length of $w_i(t)$ is equal to that of CP. Considering the ratio of CP is only 7\% of the whole symbol, the assumption is reasonable for practical wireless systems such as LTE \cite{Ref17}. It is shown in Fig. \ref{Fig2} that the simulation results match well with the theoretical analyses with varying highest derivative order \emph{N}. In Section V, with the above parameters, we will show the proposed scheme can maintain the similar spectral roll-off to traditional NC-OFDM. 

\subsection{Complexity Comparison}

%

The complexity comparison, quantified by the numbers of real additions and multiplications, among NC-OFDM, NCSP-OFDM, and the low-interference scheme is shown in Table I. Compared to other methods, the proposed low-interference scheme has notable complexity reduction. When \emph{L} is equal to or smaller than the length of CP, the complexity is significantly reduced in the transmitter. For example, when \emph{L}=$M_{\rm cp}$=144, \emph{K}=256, \emph{N}=2, 
and \emph{M}=2048, at the transmitter side, the complexity of the proposed low-interference scheme is, respectively, 47.4\% and 20.3\% of those of NC-OFDM and NCSP-OFDM.


\begin{table}[thpb]
\caption{Complexity Comparison among NC-OFDM, NCSP-OFDM and the Low-Interference Scheme in Transmitter }
\centering
\begin{tabular*}{8.85cm}{|m{1.4cm}|m{2.69cm}|m{3.45cm}|}
\hline
\bfseries Scheme & \bfseries{Number of real multiplications} & \bfseries{Number of real additions} \\
\hline
\bfseries NC-OFDM & $O(4(N + 1)K)$ & $O(2(N + 1)(2K-1))$ \\
\hline
\bfseries NCSP-OFDM & $O(8(N + 1)K)$ & $O(8(N \!+ \!1)K\!-\!4(N+1))$ \\
\hline
\bfseries Low-Interference Scheme & $O(2NK+(N+1)L)$ & $O((N+1)(2K+L+N-2))$ \\
\hline
\end{tabular*}
\end{table}

\section{Numerical Results}

 Simulations are performed in a baseband-equivalent OFDM system with \emph{K}=256, $M_{\rm cp}=144$, and 16-QAM digital modulation. The carrier frequency is 2GHz, the subcarrier spacing is $\Delta f=15$KHz, and the time-domain oversampling factor is 8. 
The PSD is evaluated by Welch's averaged periodogram method \cite{Ref19} with a 2048-sample Hanning window and 512-sample overlap. In order to show the system performance in the multi-path fading environment, LTE Extended Vehicular A (EVA) channel with 9 paths of Rayleigh fading channels \cite{Ref17} is considered. 

\begin{figure}[h]
\centering
\includegraphics[width=3.4in]{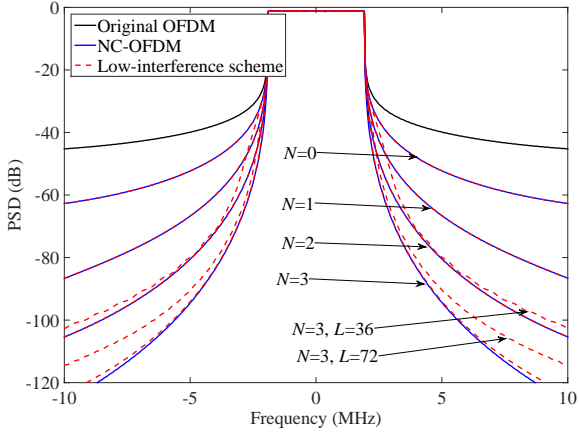} 
\DeclareGraphicsExtensions.
\caption{PSDs of NC-OFDM and the low-interference scheme with varying \emph{N} and \emph{L}.}
  \label{Fig3}
\end{figure}

Fig. \ref{Fig3} compares the PSDs between NC-OFDM and the proposed low-interference scheme with different \emph{N}. The low-interference scheme can obtain the sidelobe suppression performance similar to traditional NC-OFDM. Moreover, As \emph{L} increases, the sidelobe suppression performance of the proposed scheme is improved. For example, \emph{L} is increased from 36 to 72 and 144. In general, we show that \emph{L}=$M_{\rm cp}$=144 is a good choice for maintaining the sidelobe suppression performance as traditional NC-OFDM.

\begin{figure}[t]
\centering
\includegraphics[width=3.5in]{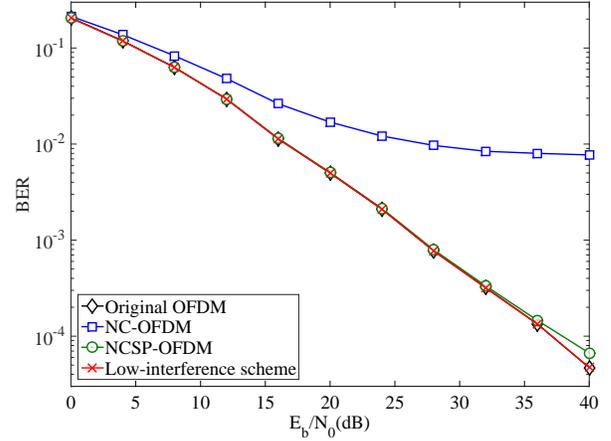}
\DeclareGraphicsExtensions.
\caption{BERs of NC-OFDM, NCSP-OFDM, and the low-interference scheme with \emph{N}=3 and \emph{L}=144 in the Rayleigh fading channel.}
  \label{Fig4}
\end{figure}

Fig. \ref{Fig4} shows the BER performance of NC-OFDM, NCSP-OFDM, and the low-interference scheme when \emph{N}=3 and \emph{L}=$M_{\rm cp}$=144  in 3GPP LTE EVA fading channel. It is shown that traditional NC-OFDM will introduce severe  interference to the transmit signal and hence to influence the BER performance, while NCSP-OFDM and our proposed low-interference scheme can both effectively mitigate the interference so as to achieve similar BER performance to OFDM, and to save the complexity for signal recovery at the receiver. However, the proposed low-interference scheme is with much lower complexity compared to other schemes according to Table I. In general, we show that the proposed scheme can make a promising tradeoff among BER performance, sidelobe suppression performance and computational complexity. 

\section{Conclusion}

In this letter, a low-interference NC-OFDM was proposed to reduce the interference and complexity as opposed to the original NC-OFDM. The main idea is to generate the time-domain low-interference smooth signal in a novel way as the linear combination of carefully designed rectangularly pulsed basis signals. Both analyses and simulation results showed that the low-interference scheme was capable of reducing the interference to a negligible extent, while maintaining similar sidelobe suppression to traditional NC-OFDM but with much lower complexity.

\end{document}